\documentclass[seceq]{ptptex}
\usepackage{wrapft}
\usepackage{bm}

\notypesetlogo  

\title{
Temporary Grand Unified Theory in Unphysical World
}

\author{
Yoshiharu \textsc{Kawamura}\footnote{E-mail: haru@azusa.shinshu-u.ac.jp}
}

\inst{
Department of Physics, Shinshu University, Matsumoto 390-8621, Japan
}

\recdate{
December 4, 2008}

\abst{
We construct grand unified models on an orbifold based on unphysical grand unification.
The reduction to the standard model or its supersymmetric one is carried out 
using a variant of Parisi-Sourlas mechanism
and nontrivial $Z_2$ parity assignment.}

\begin{document}

\maketitle

\section{Introduction}

Recently, a grand unification scenario in unphysical world has been proposed.\cite{K}
The basic idea is as follows.
Fields form multiplets under a unified gauge group
and the gauge coupling unification originates from the gauge symmetry at a high-energy scale,
as ordinary grand unified theories (GUTs).\cite{GUT,SUSYGUT}
Some members in multiples, however, are unphysical degrees of freedom, which are eliminated by
a large local symmetry, under specific field configurations.
Then no full unified symmetry is realized in the physical world.
If physical fields are the multiples of the standard model (SM) gauge group 
and survive in the low-energy scale, 
we can understand why the SM particles (and their superpartners) play an essential role in our world
and the stability of proton.\footnote{
The suppression of proton decay is a serious problem in the minimal supersymmetric GUT.\cite{Proton}}
We refer to our scenario as topological grand unification or
unphysical grand unification.

In our scenario, the question at hand is how to formulate a grand unified theory including unphysical modes.
We assume that a Lagrangian density $\mathcal{L}_U$, which possesses a large local symmetry
and a unified gauge symmetry, is dynamically and temporarily derived from a more fundamental theory.
Our goal is that we arrive at the SM or a supersymmetric SM (SSM) after the reduction of unphysical modes.
Specifically, after eliminating extra degrees of freedom including extra coordinates,
we would like to obtain the action integral of the SM or SSM 
\begin{eqnarray}
\int d^n X {\mathcal{L}}_{U} \Rightarrow \int d^4x {\mathcal{L}}_{\tiny{\rm SM}} ~~
{\rm or} ~~ \int d^4x {\mathcal{L}}_{\tiny{\rm SSM}} ,
\label{LSM}
\end{eqnarray}
where $X$ represents an extented space-time coordinate and ${\mathcal{L}}_{\tiny{\rm SM}}$ 
$({\mathcal{L}}_{\tiny{\rm SSM}})$ is the SM (an SSM) Lagrangian density.
If this program comes off, proton stability and triplet-doublet Higgs mass splitting\footnote{
There have been many interesting proposals to realize the triplet-doublet Higgs mass splitting
beyond the minimal supersymmetric GUT.\cite{sliding}$^-$\cite{discrete}} 
can be realized to be compatible with grand unification.
The unification scale $M_U$ is regarded as an energy scale on which 
the unified theory described using ${\mathcal{L}}_U$ is derived dynamically
or the reduction to the SM (or SSM) occurs.
Hence, the grand unification temporarily appears in our framework.

In this paper, we construct grand unified models on an orbifold based on topological grand unification.
The reduction to the SM or SSM is carried out 
using a variant of Parisi-Sourlas mechanism\cite{PS}
and nontrivial $Z_2$ parity assignment.

This paper is organized as follows.
In the next section, we construct simple models to realize our scenario.
Section 3 is devoted to conclusions and discussion.

\section{Unphysical grand unification on an orbifold}

We discuss models based on the $SU(5)$ or $SU(6)$ gauge group on a space-time 
including extra commuting and anticommuting dimensions. 
Models possess a large local symmetry that eliminates unphysical degrees of freedom,
and the unified symmetry becomes artificial and unphysical.
If fields have suitable $Z_2$ parities and take specific field configurations on an orbifold,
the SM or SSM is expected to appear through the dimensional reduction.

\subsection{Extended space-time}

Space-time is assumed to be factorized into a product of four-dimensional (4D) Minkowski space-time $M^4$
and the extra space $\Omega$ including commuting and     
anticommuting dimensions,
whose coordinates are denoted by $x^{\mu}$ $(\mu = 0, 1, 2, 3)$ 
and $(y^j, \theta^j, \bar{\theta}^j)$ $(j = 1, 2)$, respectively.
Notations $X^M = (x^{\mu}, y^j, \theta^j, \bar{\theta}^j)$ 
and $x^{m} = (x^{\mu}, y^j)$ are also used.
The bosonic space is the orbifold $O^2 = R/Z_2 \times R/Z_2$
and it is constructed by dividing 2D space $R \times R$ 
with two types of $Z_2$ transformations that act on each $R$ 
according to $y^1 \to -y^1$ and $y^2 \to -y^2$, respectively.
$\theta^j$ and $\bar{\theta}^j$ are four independent Grassmann numbers satisfying 
\begin{eqnarray}
({\theta}^{j})^2 = (\bar{\theta}^{j})^2 =0 , ~~ \theta^j \bar{\theta}^{j'} = - \bar{\theta}^{j'} \theta^j , ~~
\int d\theta^j \theta^{j'} = \int d\bar{\theta}^j \bar{\theta}^{j'} = \delta^{jj'} .
\label{theta}
\end{eqnarray}
$\theta^1$ and $\theta^2$ are assumed to be transformed to $\theta^1 \to -\theta^1$ and $\theta^2 \to -\theta^2$
in concert with $y^1 \to -y^1$ and $y^2 \to -y^2$, respectively.\footnote{
Note that $\bar{\theta}^j$'s are not transformed.
$\theta^j$'s and $\bar{\theta}^j$'s are not spinors.
}

The field $\Phi(x, y, \theta, \bar{\theta})$ on $M^4 \times \Omega$ is, in general, expanded as
\begin{eqnarray}
&~& \Phi(x, y, \theta, \bar{\theta}) = \phi(x, y) + \sum_{j=1, 2} \theta^j \chi^j(x, y) 
 + \sum_{j=1, 2} \bar{\theta}^j \eta^j(x, y)  
\nonumber \\
&~& ~~~~~~~~~~~~~~~~~~~ + \cdots +  \bar{\theta}^1\theta^1\bar{\theta}^2\theta^2 \varphi(x, y) .
\label{SF}
\end{eqnarray}
In our model, we treat operators with the following specific form:
\begin{eqnarray}
&~& \Psi(x, y, \theta, \bar{\theta}) = \psi(x, w) \left(= \psi(x^{\mu}, w^j)\right)
\nonumber \\
&~& ~~~~~~~~~~~~~~~ = \psi(x, y) + \sum_{j=1, 2} \bar{\theta}^j\theta^j \frac{\partial}{\partial y^j}\psi(x, y) 
+  \bar{\theta}^1\theta^1\bar{\theta}^2\theta^2 \frac{\partial^2}{\partial y^1 \partial y^2}\psi(x, y)  ,
\label{SF-2}
\end{eqnarray}
where $w^j = y^j + \bar{\theta}^j \theta^j$.
In Appendix A, we propose an unconventional idea for the origin of extended space-time and the configuration (\ref{SF-2}).
According to the idea, $w^j$'s describe an extra emergent space-time and there appears the 
accidental symmetry that leaves $w^j$ unchanged such that
\begin{eqnarray}
&~& y^j \to y'^j  = y^j  - i\xi^j (\theta^j + i \bar{\theta}^j) , 
\nonumber\\
&~& \theta^j \to \theta'^j = \theta^j +  \xi^j , ~~
\bar{\theta}^j \to \bar{\theta}'^j = \bar{\theta}^j + i \xi^j ,
\label{BRST-SUSY}
\end{eqnarray}
where $\xi^j$'s are arbitrary Grassmann numbers.
We find that the integral of $\psi(x, w)$ over $M^4 \times \Omega$ turns out to be the
4D integral over $M^4$ as
\begin{eqnarray}
&~& I_{B} \equiv \int d^4x \int_{0}^{\infty} dy^1 \int_0^{\infty} dy^2 \int d^4\theta  \psi(x, w)
\nonumber \\
&~& ~~~ = \int d^4x \int_{0}^{\infty} dy^1 \int_0^{\infty} dy^2 \int d^4\theta   
\bar{\theta}^1\theta^1\bar{\theta}^2\theta^2 \frac{\partial^2}{\partial y^1 \partial y^2}\psi(x, y)  
\nonumber \\
&~& ~~~  =  \int d^4x \psi(x, 0) \equiv I_{4D} ,
\label{4D-S}
\end{eqnarray}
where $d^4\theta \equiv d\theta^1 d\bar{\theta}^1 d\theta^2 d\bar{\theta}^2$ and
we assume that $\psi(x, y)$ vanishes at infinity ($y^1 \to \infty$ and/or $y^2 \to \infty$).
This type of dimensional reduction mechanism is regarded as a variant of the Parisi-Sourlas mechanism.\cite{PS}

\subsection{Emergent SM}

Now let us construct a model with the (unphysical) $SU(5)$ gauge symmetry almost everywhere 
on the space-time $M^4 \times \Omega$
and derive the SM through the dimensional reduction. 
Our main players are the $SU(5)$ gauge field $A_M= A_M^{\alpha} T^{\alpha}$, 
a Higgs field $h$ and matter fields $(\psi_{{\bf 1}}$, $\psi_{\bar{\bf 5}}$, $\psi_{{\bf 10}})$ 
whose representations are ${\bf 24}$, ${\bf 5}$ and (${\bf 1}$, $\bar{\bf 5}$, ${\bf 10}$), respectively.
Here, $T^{\alpha}$ are $SU(5)$ gauge generators classified into two sets
(the SM gauge generators $T^a$ and the other gauge generators $T^{\hat{a}}$):
\begin{eqnarray}
&~& T^a ~~ (a = 1, \cdots, 8, 21, \cdots, 24) ~~
\Leftrightarrow ~~ ({\bf 8}, {\bf 1})_0 + ({\bf 1}, {\bf 3})_0 + ({\bf 1}, {\bf 1})_0  , 
\nonumber \\
&~& T^{\hat{a}} ~~ (\hat{a} = 9, \cdots, 20) ~~
\Leftrightarrow ~~  ({\bf 3}, {\bf 2})_{-5/6} + (\bar{\bf 3}, {\bf 2})_{5/6} .
\label{Ta}
\end{eqnarray}
Here, the boldface numbers represent gauge quantum numbers under the SM gauge group
$G_{\tiny{\rm SM}} = SU(3)_C \times SU(2)_L \times U(1)_Y$.
$Z_2$ parities on $\Omega$ are defined using the $Z_2$ transformation properties:
\begin{eqnarray}
&~& A_{\mu}(x, -y^1, y^2, -\theta^1, \bar{\theta}^1, \theta^2, \bar{\theta}^2) 
= P_1 A_{\mu}(x, y^1, y^2, \theta^1, \bar{\theta}^1, \theta^2, \bar{\theta}^2) P_1^{-1} ,
\nonumber \\
&~& A_{5}(x, -y^1, y^2, -\theta^1, \bar{\theta}^1, \theta^2, \bar{\theta}^2) 
= - P_1 A_{5}(x, y^1, y^2, \theta^1, \bar{\theta}^1, \theta^2, \bar{\theta}^2) P_1^{-1} ,
\nonumber \\
&~& A_{6}(x, -y^1, y^2, -\theta^1, \bar{\theta}^1, \theta^2, \bar{\theta}^2) 
= P_1 A_{6}(x, y^1, y^2, \theta^1, \bar{\theta}^1, \theta^2, \bar{\theta}^2) P_1^{-1} ,
\nonumber \\
&~& A_{\theta^1}(x, -y^1, y^2, -\theta^1, \bar{\theta}^1, \theta^2, \bar{\theta}^2) 
=  - P_1 A_{\theta^1}(x, y^1, y^2, \theta^1, \bar{\theta}^1, \theta^2, \bar{\theta}^2) P_1^{-1} ,
\nonumber \\
&~& A_{\bar{\theta}^1}(x, -y^1, y^2, -\theta^1, \bar{\theta}^1, \theta^2, \bar{\theta}^2) 
=  P_1 A_{\bar{\theta}^1}(x, y^1, y^2, \theta^1, \bar{\theta}^1, \theta^2, \bar{\theta}^2) P_1^{-1} ,
\nonumber \\
&~& A_{\theta^2}(x, -y^1, y^2, -\theta^1, \bar{\theta}^1, \theta^2, \bar{\theta}^2) 
=   P_1 A_{\theta^2}(x, y^1, y^2, \theta^1, \bar{\theta}^1, \theta^2, \bar{\theta}^2) P_1^{-1} ,
\nonumber \\
&~& A_{\bar{\theta}^2}(x, -y^1, y^2, -\theta^1, \bar{\theta}^1, \theta^2, \bar{\theta}^2) 
=  P_1 A_{\bar{\theta}^2}(x, y^1, y^2, \theta^1, \bar{\theta}^1, \theta^2, \bar{\theta}^2) P_1^{-1} ,
\nonumber \\
&~& h(x, -y^1, y^2, -\theta^1, \bar{\theta}^1, \theta^2, \bar{\theta}^2) 
=  \eta_h P_1 h(x, y^1, y^2, \theta^1, \bar{\theta}^1, \theta^2, \bar{\theta}^2) ,
\label{P1} \\
&~& A_{\mu}(x, y^1, -y^2, \theta^1, \bar{\theta}^1, -\theta^2, \bar{\theta}^2) 
= P_2 A_{\mu}(x, y^1, y^2, \theta^1, \bar{\theta}^1, \theta^2, \bar{\theta}^2) P_2^{-1} ,
\nonumber \\
&~& A_{5}(x, y^1, -y^2, \theta^1, \bar{\theta}^1, -\theta^2, \bar{\theta}^2) 
=  P_2 A_{5}(x, y^1, y^2, \theta^1, \bar{\theta}^1, \theta^2, \bar{\theta}^2) P_2^{-1} ,
\nonumber \\
&~& A_{6}(x, y^1, -y^2, \theta^1, \bar{\theta}^1, -\theta^2, \bar{\theta}^2) 
= -P_2 A_{6}(x, y^1, y^2, \theta^1, \bar{\theta}^1, \theta^2, \bar{\theta}^2) P_2^{-1} ,
\nonumber \\
&~& A_{\theta^1}(x, y^1, -y^2, \theta^1, \bar{\theta}^1, -\theta^2, \bar{\theta}^2) 
=  P_2 A_{\theta^1}(x, y^1, y^2, \theta^1, \bar{\theta}^1, \theta^2, \bar{\theta}^2) P_2^{-1} ,
\nonumber \\
&~& A_{\bar{\theta}^1}(x, y^1, -y^2, \theta^1, \bar{\theta}^1, -\theta^2, \bar{\theta}^2) 
=  P_2 A_{\bar{\theta}^1}(x, y^1, y^2, \theta^1, \bar{\theta}^1, \theta^2, \bar{\theta}^2) P_2^{-1} ,
\nonumber \\
&~& A_{\theta^2}(x, y^1, -y^2, \theta^1, \bar{\theta}^1, -\theta^2, \bar{\theta}^2) 
= -P_2 A_{\theta^2}(x, y^1, y^2, \theta^1, \bar{\theta}^1, \theta^2, \bar{\theta}^2) P_2^{-1} ,
\nonumber \\
&~& A_{\bar{\theta}^2}(x, y^1, -y^2, \theta^1, \bar{\theta}^1, -\theta^2, \bar{\theta}^2) 
= P_2 A_{\bar{\theta}^2}(x, y^1, y^2, \theta^1, \bar{\theta}^1, \theta^2, \bar{\theta}^2) P_2^{-1} ,
\nonumber \\
&~& h(x, y^1, -y^2, \theta^1, \bar{\theta}^1, -\theta^2, \bar{\theta}^2) 
=  \eta'_h P_2 h(x, y^1, y^2, \theta^1, \bar{\theta}^1, \theta^2, \bar{\theta}^2) ,
\label{P2} 
\end{eqnarray}
where $P_1$ and $P_2$ are $5 \times 5$ matrices and $P_1^2 = P_2^2 = I_{5 \times 5}$ 
($I_{5 \times 5}$ is a $5 \times 5$ unit matrix) by definition.
The $\eta_{h}$ and $\eta'_{h}$ are intrinsic parities of $h$.
The transformation laws (\ref{P1}) and (\ref{P2}) are consistent with the gauge covariance of the derivative 
$D_{M} = \partial_{M} + ig_{U} A_{M}$ where $g_U$ is a unified gauge coupling constant.

Let us assume that the gauge bosons and the Higgs boson are present in the bulk
whose field configurations are dynamically fixed (up to a freedom of 
residual gauge symmetry) as $A_M(x, w)$ and $h(x, w)$ 
with the following conditions:
\begin{eqnarray}
F_{M\theta^j}(x,w) = F_{M\bar{\theta}^j}(x,w) = D_{\theta^j} h(x,w) = D_{\bar{\theta}^j} h(x,w)  = 0 
\label{cond}
\end{eqnarray}
on the scale $M_U$ by a more fundamental theory.
Here, $F_{MN}$ is the field strength of $A_{M}$. 
Furthermore, we assume that matter fermions are localized at the origin on $\Omega$
as 4D left-handed chiral ones $(\psi_{{\bf 1}}(x)$, 
$\psi_{\bar{\bf 5}}(x)$, $\psi_{{\bf 10}}(x))$.
The gauge-transformed configurations are equivalent to the original ones
with a restricted gauge transformation function, $U(x,w)=\exp(i\xi^{\alpha}(x,w)T^{\alpha})$.
The gauge transformations for gauge bosons and Higgs boson are given by
\begin{eqnarray}
&~& A_M(x,w) \to A'_M(x,w) 
= U(x,w) A_M(x,w) U^{-1}(x,w) - \frac{i}{g_U} U(x,w) \partial_M U^{-1}(x,w) ,
\nonumber \\
&~& h(x,w) \to h'(x,w) = U(x,w) h(x,w) .
\label{gauge-inv}
\end{eqnarray}
The gauge transformations for matter fields are given with $U(x, 0)$.
The $Z_2$ transformation properties become
\begin{eqnarray}
&~& A_{\mu}(x, -w^1, w^2) = P_1 A_{\mu}(x, w^1, w^2) P_1^{-1} , ~
A_{5}(x, -w^1, w^2) = - P_1 A_{5}(x, w^1, w^2) P_1^{-1} ,
\nonumber \\
&~& A_{6}(x, -w^1, w^2) = P_1 A_{6}(x, w^1, w^2) P_1^{-1} , ~ h(x, -w^1, w^2) =  \eta_h P_1 h(x, w^1, w^2) ,
\label{P1-w} \\
&~& A_{\mu}(x, w^1, -w^2) = P_2 A_{\mu}(x, w^1, w^2) P_2^{-1} , ~
A_{5}(x, w^1, -w^2) =  P_2 A_{5}(x, w^1, w^2) P_2^{-1} ,
\nonumber \\
&~& A_{6}(x, w^1, -w^2) = -P_2 A_{6}(x, w^1, w^2) P_2^{-1} , ~
h(x, w^1, -w^2) =  \eta'_h P_2 h(x, w^1, w^2) .
\label{P2-w} 
\end{eqnarray}

The gauge invariant action integral on $M^4 \times \Omega$ is given by
\begin{eqnarray}
&~& S_{U} = \int d^4x \int_{0}^{\infty} dy^1 \int_0^{\infty} dy^2 \int d^4\theta
\left(\mathcal{L}_{B}(x, w) + \delta^2(y) \delta^2(\theta) \delta^2(\bar{\theta}) \mathcal{L}_{b}(x, w)\right) ,
\nonumber \\
&~& ~ \mathcal{L}_{B}(x, w) =  -\frac{1}{2}{\rm tr}F^{mn}(x,w)F_{mn}(x,w) 
+ (D^{m} h(x,w))^{\dagger} (D_{m} h(x,w)) -V(h(x,w))
\nonumber \\
&~& ~ \mathcal{L}_{b}(x, w) = \sum_{\rm 3~families} 
\left(i\bar{\psi}_{\bar{\bf 5}}(x) \gamma^{\mu} D_{\mu} \psi_{\bar{\bf 5}}(x)
+ i\bar{\psi}_{{\bf 10}}(x) \gamma^{\mu} D_{\mu} \psi_{{\bf 10}}(x)\right. 
\nonumber \\
&~& ~~~~~~~~~~~~  \left. + f_{U} h(x,w) {\psi}_{{\bf 10}}(x) {\psi}_{{\bf 10}}(x) 
 + f_{D} h^{*}(x,w) {\psi}_{{\bf 10}}(x) \psi_{\bar{\bf 5}}(x) + {\rm h.c.}\right) ,
\label{S-SU5}
\end{eqnarray}
where $V(h(x,w))$ is a Higgs potential, $f_U$ and $f_D$ are Yukawa coupling matrices
and h.c. denotes the hermitian conjugate of the former terms.
Here and hereafter, we omit the gauge singlet neutrinos for simplicity.

After the dimensional reduction, we derive the 4D action integral
\begin{eqnarray}
&~& S_{U} = \int d^4x {\mathcal{L}}(x, 0) \equiv S_{4D} ,
\nonumber \\
&~&   {\mathcal{L}}(x, 0)  =  -\frac{1}{2}{\rm tr}F^{mn}(x,0)F_{mn}(x,0) 
+ (D^{m} h(x,0))^{\dagger} (D_{m} h(x,0)) -V(h(x,0))
\nonumber \\
&~& ~~~~~~~~~~~~~ + \sum_{\rm 3~families} \left(i \bar{\psi}_{\bar{\bf 5}}(x) \gamma^{\mu} D_{\mu} \psi_{\bar{\bf 5}}(x)
+ i \bar{\psi}_{{\bf 10}}(x) \gamma^{\mu} D_{\mu} \psi_{{\bf 10}}(x)\right. 
\nonumber \\
&~& ~~~~~~~~~~~~ \left. + f_{U} h(x,0) {\psi}_{{\bf 10}}(x) {\psi}_{{\bf 10}}(x) 
 + f_{D} h^{*}(x,0) {\psi}_{{\bf 10}}(x) \psi_{\bar{\bf 5}}(x) + {\rm h.c.}\right) ,
\label{S-4D}
\end{eqnarray}
where we assume that ${\mathcal{L}}_{B}(x, y)$ vanishes at infinity ($y^1 \to \infty$ and/or $y^2 \to \infty$).

As shown by (\ref{S-4D}), the field configurations around the origin on $\Omega$ determine the theory on $M^4$.
Specifically, if only SM gauge bosons $A^{a}_{\mu}$ and weak Higgs boson $h_W$ (in addition to
SM fermions) survive at the origin on $\Omega$, i.e.,
\begin{eqnarray}
A^{a}_{\mu}(x,0) \ne 0 , ~~ h_W(x,0) \ne 0 
\label{conf}
\end{eqnarray}
and extra particles and operators relating extra dimensions vanish there, i.e.,
\begin{eqnarray}
&~& A^{\hat{a}}_{\mu}(x,0) = 0 , ~~ h_C(x,0) = 0 ,
\nonumber \\
&~& F_{5\mu}(x,0) = F_{6\mu}(x,0) = F_{56}(x,0) = 0 , ~~
D_5 h(x,0) = D_6 h(x,0) = 0 ,
\label{conf2}
\end{eqnarray}
the action integral $S_{4D}$ turns out to be the SM one without the QCD $\theta$ term.
The absence of the strong $CP$-violating term is due to the fact that there is no term
in the action integral (\ref{S-SU5}) to generate it.\footnote{
This reasoning is the same as that in Ref. \citen{2T}.
In Ref. \citen{2T}, the emergence of SM (or its dual theory) from higher dimensional theory 
has been discussed in the framework of 2T-physics.
Their idea is similar to ours, but the starting point and guiding principle are different from ours.}
Hence, the strong {\it CP} problem can be solved if the value of ${\rm argdet}(M_u M_d)$ is sufficiently suppressed
where $M_u$ and $M_d$ are mass matrices of the up- and down-type quarks, respectively.

The disappearance of extra fields ($A^{\hat{a}}_{\mu}(x,w)$, $h_C(x,w)$)
and operators ($F_{5\mu}(x,w)$, $F_{6\mu}(x,w)$, $F^{a}_{56}(x,w)$, 
$D_5 h(x,w)$, $D_6 h(x,w)$) at the origin on $\Omega$ is realized using the 
$Z_2$ parity assignment $P_1=P_2={\rm diag}(-1,-1,-1,1,1)$.\footnote{
If we take the parity assignment $P_1 = P_2 = I_{5 \times 5}$, we derive the relation
$A_5(x, 0) = A_6(x, 0)$ $= F_{5\mu}(x, 0) = F_{6\mu}(x, 0)$ $= F_{56}(x, 0) = D_5 h(x, 0)$ $= D_6 h(x, 0) = 0$.
Then the minimal $SU(5)$ GUT (without an adjoint Higgs boson) is obtained 
with the intrinsic parities $\eta_h = \eta'_h = 1$ and appropriate field configurations.}
This originates from the fact that fields and operators with odd $Z_2$ parity vanish at the origin on $\Omega$.
The $Z_2$ parities are given with the intrinsic ones $\eta_h = \eta'_h = 1$ in Table \ref{t1}.
\begin{table}
\caption{$Z_2$ Parity.}
\label{t1}
\begin{center}
\begin{tabular}{c|c|c} \hline \hline
Fields & Quantum numbers & Parity \\ \hline
$A_{\mu}^a$ & $({\bf 8}, {\bf 1})_0 + ({\bf 1}, {\bf 3})_0 + ({\bf 1}, {\bf 1})_0$ & $(+, +)$ \\
$A_{\mu}^{\hat{a}}$ & $({\bf 3}, {\bf 2})_{-5/6} + (\bar{\bf 3}, {\bf 2})_{5/6}$ & $(-, -)$ \\ \hline
$A_{5}^a$ & $({\bf 8}, {\bf 1})_0 + ({\bf 1}, {\bf 3})_0 + ({\bf 1}, {\bf 1})_0$ & $(-, +)$ \\
$A_{5}^{\hat{a}}$ & $({\bf 3}, {\bf 2})_{-5/6} + (\bar{\bf 3}, {\bf 2})_{5/6}$ & $(+, -)$ \\ \hline
$A_{6}^a$ & $({\bf 8}, {\bf 1})_0 + ({\bf 1}, {\bf 3})_0 + ({\bf 1}, {\bf 1})_0$ & $(+, -)$ \\
$A_{6}^{\hat{a}}$ & $({\bf 3}, {\bf 2})_{-5/6} + (\bar{\bf 3}, {\bf 2})_{5/6}$ & $(-, +)$ \\ \hline
$h_C$ & $({\bf 3}, {\bf 1})_{-1/3}$ & $(-, -)$ \\
$h_W$ & $({\bf 1}, {\bf 2})_{1/2}$ & $(+, +)$ \\ \hline
\end{tabular}
\end{center}
\end{table}
The gauge invariance is reduced to the SM one at the origin on $\Omega$
and the splitting between doublet Higgs boson and triplet one is realized, 
as heterotic string models\cite{SST} and grand unified theories on an orbifold.\cite{orb}
This is because the parity assignment does not respect $SU(5)$ gauge symmetry,
as we see from the relations for the gauge generators,
\begin{eqnarray}
P_1 T^a P_1 = T^a , ~~ P_1 T^{\hat{a}} P_1 = - T^{\hat{a}} , ~~ P_2 T^a P_2 = T^a , ~~ P_2 T^{\hat{a}} P_2 = - T^{\hat{a}} . 
\label{relations}
\end{eqnarray}
The gauge transformation function compatible with the $Z_2$ parity assignment
is constrained such that
\begin{eqnarray}
&~& \xi^{a}(x,0) \ne 0 , ~~ \xi^{\hat{a}}(x,0) = 0 .
\label{xi}
\end{eqnarray}
Note that $SU(5)$ gauge symmetry exists where $\xi^{a}(x,w) \ne 0$ and $\xi^{\hat{a}}(x,w) \ne 0$.
The $F^{\hat{a}}_{56}(x,w)$ can survive at the origin of $\Omega$, but it would be harmless
because $A_{5}$ and $A_6$ neither locally couple to SM particles nor have dynamical degrees of freedom.

\subsection{Emergent supersymmetric SM}

The SM has been established as an effective theory on a weak scale, but it suffers from several problems.
Typical ones are the naturalness problem 
and the dark matter problem. 
These problems are solved with the introduction of space-time supersymmetry.
As a bonus, the gauge coupling unification occurs on the basis of the minimal supersymmetric SM (MSSM)
with a big desert hypothesis between the TeV scale and the grand unification scale.\cite{GCU}

Let us discuss the space-time supersymmetric extension of our model. 
Our main players are the $N=2$ supersymmetric vector multiplet on $M^4 \times \Omega$, i.e.,
the $SU(6)$ vector multiplet $V=(A_{\mu}, \lambda_1)$ and three chiral multiplets
$\Sigma_5 = (\sigma_1 + i A_5, \lambda_2)$, $\Sigma_6 = (\sigma_2 + i A_6, \lambda_3)$,  
$\Sigma = (\sigma + i \rho, \lambda_4)$ in terms of $N=1$ supersymmetry in 4D, 
and $SU(5)$ matter chiral superfields $\Psi_{\bm{n}} = (\tilde{\psi}_{\bm{n}}, \psi_{\bm{n}})$
$({\bm{n}}$ $= {\bf 1}$, $\bar{\bf 5}$, ${\bf 10})$
located at the origin on $\Omega$.
Here, $\lambda_s$ $(s = 1 \sim 4)$ are gauginos
and $\tilde{\psi}_{\bm{n}}$s are sfermions (superpartners of matter fields).

We assume that the following action appears at $M_U$,
\begin{eqnarray}
&~& S_{U}^{\tiny{\rm SUSY}} = \int d^4x \int_{0}^{\infty} dy^1 \int_0^{\infty} dy^2 \int d^4\theta
({\mathcal{L}}_{B}(x, w) 
\nonumber \\
&~& ~~~~~~~~~~~~~~~~~~~~~~~~~~~~~~~~~ + \delta^2(y) \delta^2(\theta) \delta^2(\bar{\theta}) \mathcal{L}_{b}(x,w)),
\label{S-SUSY}
\end{eqnarray}
where ${\mathcal{L}}_{B}(x, w)$ is the $SU(6)$ gauge invariant Lagrangian density 
of $N=2$ supersymmetric vector multiplet.\footnote{
See Ref.~\citen{N4D4} for $N=4$ super Yang-Mills theory on 4D space-time
and Ref.~\citen{N2D6} for $N=2$ super Yang-Mills theory on 6D space-time.}
Here, $\mathcal{L}_{b}(x,w)$ is the Lagrangian density given by
\begin{eqnarray}
&~&  {\mathcal{L}}_{b}(x, w)  =  
 \sum_{\rm 3~families} (i \bar{\psi}_{\bar{\bf 5}}(x) \gamma^{\mu} D_{\mu} \psi_{\bar{\bf 5}}(x)
 + i \bar{\psi}_{{\bf 10}}(x) \gamma^{\mu} D_{\mu} \psi_{{\bf 10}}(x)
\nonumber \\
&~& ~~~ + |D_{\mu} \tilde{\psi}_{\bar{\bf 5}}(x)|^2 + |D_{\mu} \tilde{\psi}_{\bf 10}(x)|^2
\nonumber \\
&~& ~~~ + f_{U} h(x,w) {\psi}_{{\bf 10}}(x) {\psi}_{{\bf 10}}(x) 
+ f_{D} h^{*}(x,w) {\psi}_{{\bf 10}}(x) \psi_{\bar{\bf 5}}(x) + {\rm h.c.} + \cdots) ,
\label{S-8D-SUSY}
\end{eqnarray}
where $h(x,w)$ and $h^{*}(x,w)$ are Higgs bosons, which originate from $\Sigma$, 
whose representations are ${\bf 5}$ and $\bar{\bf 5}$ under $SU(5)$. 
${\mathcal{L}}_{b}(x, w)$ possesses only $SU(5)$ gauge invariance
because bulk fields are $SU(6)$ multiplets and matter fields form $SO(10)$ multiplets.
Here, we require that the symmetry of ${\mathcal{L}}_{b}(x, w)$ is determined 
by the intersection of the gauge group in the bulk and the largest group, which localized fields
form the full multiplet.
This leads to the gauge coupling unification and the mass unification among members
in each multiplet.\footnote{
If we take a standpoint that there are localized terms allowed by the symmetry (including $G_{\tiny{\rm SM}}$) 
in the theory after the reduction, the gauge coupling unification and the mass unification do not necessarily hold.}
Note that matter fields do not locally couple to any members in $\Sigma_5$ and $\Sigma_6$ 
from the higher-dimensional gauge invariance. 
The MSSM fields are $A^a_{\mu}$, $\lambda^a_1$, $h_{W1}$, $h_{W2}$, $\tilde{h}_{W1}$, $\tilde{h}_{W2}$,
$\tilde{\psi}_{\bm{n}}$ and $\psi_{\bm{n}}$.
If ${\mathcal{L}}_B(x, w)$ vanishes at infinity on $\Omega$ and only the MSSM particles survive at the origin on $\Omega$,
the MSSM action without the $\mu$ term can be derived after the dimensional reduction.
The $\mu$ term is forbidden by the $R$ symmetry.
The $\mu$ term with a suitable magnitude could be generated upon the breakdown of space-time SUSY, e.g.,
through Guidice-Masiero mechanism.\cite{GM}

Now the problem is whether the disappearance of extra particles at the origin on $\Omega$ 
can be realized using a suitable $Z_2$ parity assignment? 
If we take $P_1={\rm diag}(1,1,1,1,1,-1)$ and $P_2={\rm diag}(-1,-1,-1,1,1,-1)$,
$Z_2$ parities of fields are given in Table \ref{t2}.
\begin{table}
\caption{$Z_2$ Parity in SUSY model.}
\label{t2}
\begin{center}
\begin{tabular}{c|c|c} \hline \hline
Fields & Quantum numbers & Parity \\ \hline
$A_{\mu}^{a}$, $\lambda^a_1$ & $({\bf 8}, {\bf 1})_{0} + ({\bf 1}, {\bf 3})_{0} 
+ ({\bf 1}, {\bf 1})_{0} + ({\bf 1}, {\bf 1})_{0}$ & $(+, +)$ \\
$A_{\mu}^{\hat{a}}$, $\lambda^{\hat{a}}_1$  & $({\bf 3}, {\bf 2})_{-5/6} + (\bar{\bf 3}, {\bf 2})_{5/6}$ & $(+, -)$ \\ 
$A_{\mu}^b$, $\lambda^b_1$ & $({\bf 3}, {\bf 1})_{-1/3} + (\bar{\bf 3}, {\bf 1})_{1/3}$ & $(-, +)$ \\
$A_{\mu}^{\hat{b}}$, $\lambda^{\hat{b}}_1$  & $({\bf 1}, {\bf 2})_{1/2} + ({\bf 1}, {\bf 2})_{-1/2}$ & $(-, -)$ \\ \hline
$\sigma_1^{a} + iA_{5}^{a}$, $\lambda^a_2$ & $({\bf 8}, {\bf 1})_{0} + ({\bf 1}, {\bf 3})_{0} 
+ ({\bf 1}, {\bf 1})_{0} + ({\bf 1}, {\bf 1})_{0}$ & $(-, +)$ \\
$\sigma_1^{\hat{a}} + iA_{5}^{\hat{a}}$, $\lambda^{\hat{a}}_2$  
& $({\bf 3}, {\bf 2})_{-5/6} + (\bar{\bf 3}, {\bf 2})_{5/6}$ & $(-, -)$ \\ 
$\sigma_1^{b} + iA_{5}^{b}$, $\lambda^b_2$ & $({\bf 3}, {\bf 1})_{-1/3} + (\bar{\bf 3}, {\bf 1})_{1/3}$ & $(+, +)$ \\
$\sigma_1^{\hat{b}} + iA_{5}^{\hat{b}}$, $\lambda^{\hat{b}}_2$  
& $({\bf 1}, {\bf 2})_{1/2} + ({\bf 1}, {\bf 2})_{-1/2}$ & $(+, -)$ \\ \hline
$\sigma_2^{a} + iA_{6}^{a}$, $\lambda^a_3$ & $({\bf 8}, {\bf 1})_{0} + ({\bf 1}, {\bf 3})_{0} 
+ ({\bf 1}, {\bf 1})_{0} + ({\bf 1}, {\bf 1})_{0}$ & $(+, -)$ \\
$\sigma_2^{\hat{a}} + iA_{6}^{\hat{a}}$, $\lambda^{\hat{a}}_3$  
& $({\bf 3}, {\bf 2})_{-5/6} + (\bar{\bf 3}, {\bf 2})_{5/6}$ & $(+, +)$ \\ 
$\sigma_2^{b} + iA_{6}^{b}$, $\lambda^b_3$ & $({\bf 3}, {\bf 1})_{-1/3} + (\bar{\bf 3}, {\bf 1})_{1/3}$ & $(-, -)$ \\
$\sigma_2^{\hat{b}} + iA_{6}^{\hat{b}}$, $\lambda^{\hat{b}}_3$  
& $({\bf 1}, {\bf 2})_{1/2} + ({\bf 1}, {\bf 2})_{-1/2}$ & $(-, +)$ \\ \hline
$\sigma^{a} + i\rho^{a}$, $\lambda^a_4$ & $({\bf 8}, {\bf 1})_{0} + ({\bf 1}, {\bf 3})_{0} 
+ ({\bf 1}, {\bf 1})_{0} + ({\bf 1}, {\bf 1})_{0}$ & $(-, -)$ \\
$\sigma^{\hat{a}} + i\rho^{\hat{a}}$, $\lambda^{\hat{a}}_4$  
& $({\bf 3}, {\bf 2})_{-5/6} + (\bar{\bf 3}, {\bf 2})_{5/6}$ & $(-, +)$ \\ 
$\sigma^{b} + i\rho^{b}$, $\lambda^b_4$ & $({\bf 3}, {\bf 1})_{-1/3} + (\bar{\bf 3}, {\bf 1})_{1/3}$ & $(+, -)$ \\
$\sigma^{\hat{b}} + i\rho^{\hat{b}}$, $\lambda^{\hat{b}}_4$  
& $({\bf 1}, {\bf 2})_{1/2} + ({\bf 1}, {\bf 2})_{-1/2}$ & $(+, +)$ \\ \hline
\end{tabular}
\end{center}
\end{table}
We find that ($\sigma_1^{b} + iA_{5}^{b}$, $\lambda^b_2$) 
and ($\sigma_2^{\hat{a}} + iA_{6}^{\hat{a}}$, $\lambda^{\hat{a}}_3$), 
in addition to the MSSM gauge multiplets ($A_{\mu}^{a}$, $\lambda^a_1$) 
and Higgs multiplets ($\sigma^{\hat{b}} + i\rho^{\hat{b}}$, $\lambda^{\hat{b}}_4$),
have even $Z_2$ parities and could survive at the origin on $\Omega$.
The proton stability, however, is still gauranteed
because they do not couple to matter fields from the higher-dimensional gauge invariance.
The realization of gauge coupling unification is a challenging problem in this setup.

\section{Conclusions and discussion}

We have constructed unified models with the $SU(5)$ or $SU(6)$ gauge group 
on the orbifold $R/Z_2 \times R/Z_2$ based on unphysical grand unification.
Dangerous colored particles are eliminated
and the reduction to the SM or an SSM can be carried out using a variant of Parisi-Sourlas mechanism,
under nontrivial $Z_2$ parity assignment and additional field configurations.
{\it The root of all evils must go back to Pandora's box by itself.}
There are open questions for our models.
Is it possible to select $Z_2$ parities and
arrange field configurations to realize the SM or SSM?
If possible, how are they fixed and are they stable against radiative corrections?
Is it possible to derive our grand unified theories from a more fundamental theory such as
superstring theory?

Since extra particles are unphysical in topological grand unification, 
it can be checked using the SSM particles features and their effects.
Concretely, there are no threshold corrections to modify the running of parameters 
due to renormalization group effects if no physical particles with SM gauge quantum numbers 
appear around the unification $M_U$. 
Then unification of parameters such as gauge couplings or some particle masses can occur at $M_U$ 
if there are no sizable localized terms that violate a unified symmetry.
Hence, gaugino masses and sfermion masses can be useful probes of our scenario
by checking whether those masses agree at $M_U$ among members 
in each multiplet of unified gauge group.\cite{KMY}

We have proposed an unconventional idea 
that the extended space-time and the configuration (\ref{SF-2}) emerge after the breakdown of 
symmetry (\ref{symmetry}) following a topology change.
The crucial assumption is that the topology change occurs
with the formation of bound states made of ghost variables,
and ghost variables survive as fluctuations of bound states.
It is intriguing to study the dynamics related to the topology change. 
In our scenario, $M_U$ is considered to be the scale 
where the topology change occurs.
Models have been constructed on the flat space-time without considering the gravity,
assuming that $M_U$ is different from the Planck scale $M_P$.
Gravitational effects, however, must be considered if $M_U$ is near $M_P$.
Study on elementary particle physics using a topological symmetry\footnote{
An $OSp(4/2)$ supersymmetric model and a deformation of topological field theory
were studied to explore the quark confinement mechanism in QCD.\cite{HK,Kondo}}
would be attractive and it is worth exploring a high-energy theory on the basis of our insight.

\section*{Acknowledgements}
This work was supported in part by scientific grants from the Ministry of Education, Culture, Sports, Science and Technology
under Grant Nos.~18204024 and 18540259.

\appendix
\section{Origin of Extended Space-Time Structure}

We provide an idea for the origin of extended space-time and the configuration (\ref{SF-2})
on the basis of assumptions.

We assume that {\it the space-time manifold, whose coordinates are described 
using bosonic variables $x^a$ $(a = 1, \cdots, 4)$, is compact and orientable at the microscopic level,
and there is a symmetry under which $x^a$ is transformed into ghost ones $\theta^a$}.
This assumption is based on the following observation.
Our space-time is macroscopically four-dimensional Minkowski (puedo-Riemannian) space
in the absence (presence) of gravity.
It is widely considered that the structure of space-time and/or quantum field theory 
must be modified below the Planck length $l_{P}$
because quantum gravitational effects are no longer negligible and 
are out of control using a conventional quantum field theory.
It is possible to handle quantum gravitational effects
with the appearance of a large local symmetry such as topological symmetry,
where the graviton turns out to be unphysical in its unbroken phase.\cite{TFT}

Let the symmetry be generated by the operator with nilpotency $Q = \theta^a p_a$ and 
the transformation be given by
\begin{eqnarray}
\hat{\delta} x^a = \theta^a , ~~ \hat{\delta} \theta^a = 0 , ~~
\hat{\delta} \eta_a = p_a , ~~ \hat{\delta} p_a = 0 ,
\label{symmetry}
\end{eqnarray}
where a Grassmann parameter is omitted and 
$p_a$ and $\eta_a$ are the canonical conjugate momenta of $x^a$ and $\theta^a$, respectively.
Fields are functions of $x^{a}$ and $\theta^{a}$, and
there are no local observables in the unbroken phase of (\ref{symmetry}).

Furthermore, we assume that {\it the symmetry (\ref{symmetry}) is broken down (at a larger scale than $l_P$) 
following the topology change with the formation of bound states made of ghost variables,
and ghost variables survive as fluctuations of bound states}.
This assumption is based on the following expectation.
It is necessary to break the symmetry (\ref{symmetry}) to create locally physical degrees of freedom.
It is, in general, difficult to break the topological symmetry spontaneously.\cite{TFT2}
Topology change can be a seed of local physics.

Let ghost variables form two bound states $y^1=[\bar{\theta}^1 \theta^1]$ and $y^2=[\bar{\theta}^2 \theta^2]$, 
and we assume that these states take any real values on some region 
such as $R/Z_2 \times R/Z_2$ owing to the existence of flat directions.
Here, we write $\bar{\theta}^1$ and $\bar{\theta}^2$ in place of $i \theta^3$ and $i \theta^4$, respectively.
The $y^1$ and $y^2$ are treated as hermitian bosonic variables with the ghost number two.\footnote{
We assume that the ghost number symmetry is also broken down following the topology change.}
Other pairings such as $i[\theta^1 \theta^2]$, $i[\theta^1 \theta^4]$, $i[\theta^2 \theta^3]$ 
and $i[\theta^3 \theta^4]$ are assumed to be forbidden to form.
If the structure of space-time changes from Euclidean (Riemannian) space to Minkowski (puedo-Riemannian) space,
i.e., space coordinates $(x^4, \theta^4)$ change into time coordinates,
$y^1$ ($y^2$) is regarded as an emergent extra space (time) coordinate.\footnote{Unless ghost variables 
are treated as physical ones, 
$(4+2)$-dimensional space-time  
occurs and can offer the framework of $2T$ physics.\cite{2T}}
If ghost variables survive as fluctuations of bound states,
coordinate variables are given by
\begin{eqnarray}
x^0, ~ x^1, ~ x^2, ~ x^3; ~
w^1 \equiv y^1 + \bar{\theta}^1 \theta^1 , ~ 
w^2 \equiv y^2 + \bar{\theta}^2 \theta^2 ,
\label{w}
\end{eqnarray}
where $x^0$ is used in place of $x^4$ and $\bar{\theta}^1 \theta^1$ and $\bar{\theta}^2 \theta^2$ are fluctuations
of $y^1$ and $y^2$, respectively.
In this manner, the extended space-time $M^4 \times \Omega$ originates macroscopically
and fields are introduced as functions of $x^{\mu}$ ($\mu = 0, 1, 2, 3$) and $w^j$ ($j=1,2$).

\end{document}